\documentclass[11pt,twoside]{article}
\usepackage{thmt_rev}
\usepackage[dvips]{graphics}
\usepackage{epsf}
\usepackage{style}
\usepackage{times}
\runninghead{T. Tsukahara et al.}

\title{DNS of Heat Transfer in a Transitional Channel Flow\\
Accompanied by a Turbulent Puff-like Structure}

\author{Takahiro TSUKAHARA$^*$, Kaoru IWAMOTO, Hiroshi KAWAMURA, Tetsuaki TAKEDA\\ \vskip -0.5em
{\normalsize \noindent $^*$Department of Mechanical Engineering, Tokyo University of Science. E-mail: {\it tsuka@rs.tus.ac.jp}}}

\begin{document}

\maketitle

\begin{abstract}   
Direct numerical simulations of turbulent heat transfer in fully-developed channel flows have been performed in a range of friction Reynolds number between 60 and 180, based on the friction velocity and the channel half width $\delta$, with emphasis on a puff-like structure, large-scale spatial intermittency. 
For the Reynolds numbers lower than 80 with a large computational domain of $51.2 \times 2 \times 22.5$, the turbulent puff was observed and its significant influences on the momentum and heat transports were found. 
The spatial structure of the equilibrium puff, or the localized turbulence, was examined with taking account of two different thermal boundary conditions: the uniform heat-flux heating and the constant temperature difference between the walls.
It was revealed that there existed a localized strong turbulent region in the form of an oblique band, along which a spanwise secondary flow was induced.
In consequence, at the present lowest Reynolds number as low as 60, the flow remained turbulent and the larger Nusselt numbers than those without puff was obtained by the presence of puff. 
\end{abstract}

\section{Introduction}
Heat transfer at low Reynolds numbers in turbulent/transitional channel flow is of practical importance. 
Since the process of relaminarization is also important in the field of both engineering applications and fundamental flow physics, the transition from turbulence to laminar has been studied experimentally by a number of researchers. 

In these days, the direct numerical simulation (DNS) is a powerful tool to study the turbulent heat transfer in a channel.
The first such DNS was made by Kim and Moin \cite{Kim89} with an assumption of uniform heat generation at $Re_\tau=180$, where $Re_\tau$ is based on the friction velocity $u_\tau$ and the channel half width $\delta$. 
Lyons \etal \cite{Lyons91a,Lyons91b} performed the simulation at $Re_\tau=150$ with the constant temperature difference boundary condition.
At the same Reynolds number, Kasagi \etal \cite{Kasagi92} executed DNS under the condition of uniformly heated walls.
Later, the Reynolds- and Prandtl-number dependencies have been investigated by many research groups (e.g., Kawamura \etal \cite{Kawa98,Kawa00}).
On the other hand, not much work has been done on DNS of heat transfer in a turbulent/transitional channel flow for a lower Reynolds number than $Re_\tau=150$. 
This might be because it would require a mcuh larger computational domain in order to capture an elongated streaky structure that is scaled in the wall units.
As for the transitional channel flow without a scalar transport, Iida and Nagano \cite{Iida98} carried out the DNS at $Re_\tau=60$--$100$ to investigate the mechanisms of laminarization.
The authors' group \cite{Tsukahara05} showed that an isolated-turbulent structure (termed `puff-like structure' in this paper) was observed at $Re_\tau=80$ by expanding the domain.
The structure was found to be similar to the `equilibrium turbulent puff' in transitional pipe flows reported by Wygnanski and Champagne \cite{Wygnanski73}.

Since most channel flows in various engineering systems can undergo laminar-to-turbulent transition below the critical Reynolds number given by the linear stability theory, the subcritical transition in a channel flow has also been studied in many works. 
An early experimental study on the transitional channel flow was made by Davies \& White \cite{Davies28}. 
Comparison of the existing experimental results indicates wide variation in the transitional Reynolds numbers.
Patel \& Head \cite{Patel69} obtained the transition Reynolds number of about $Re_{\rm c}=1035$ with high-intensity disturbance in an inlet flow.
Carlson \etal \cite{Carlson82} made a flow-visualization study of turbulent spots with a smooth inlet in order to achieve a low-turbulence background flow.
Their observations indicated that natural turbulent spots appeared spontaneously, leading to a transition at $Re_{\rm c}$ (defined in \Tref{tab:condition}) slightly above 1000 as in Patel \& Head \cite{Patel69}.
These transitional Reynolds numbers are less than 20\% 
of the critical Reynolds number ($Re_{\rm c}=5772$) by the (two-dimensional) linear instability analysis \cite{Orszag71}.
Later, theoretical results of Orszag's group \cite{Orszag80} inferred a transitional Reynolds number of about 1000, when finite-amplitude three-dimensional disturbance was considered.

These studies show the importance of nonlinear effects and of three-dimensional secondary instability associated with transitional structures, i.e. spatially localized turbulence such as spots and puffs. 
Turbulent spots in a laminar channel flow have been studied extensively: turbulence characteristics inside a spot and its maintenance of turbulence were investigated in experiments by Klingmann \& Alfredsson \cite{Klingmann90} and DNS by Henningson \& Kim \cite{Henningson91}.
The structure of puffs in a transitional pipe flow was studied by Wygnanski's group \cite{Wygnanski73,Wygnanski75} and was found to be different from the structure of the fully-developed turbulent flow.
In the transitional channel flow, the puff-like structure first observed by Tsukahara \etal \cite{Tsukahara05}, and this transitional structure must play a role in sustaining turbulence.
Few numerical studies had focused on the effect of the transitional structure in a channel flow at turbulent-laminar transition range.
Therefore, it is necessary to know a mechanism of scalar transport in the puff-like structure with respect to enhancement of heat transfer.

\begin{table}[t]
 \caption{Reynolds numbers and domain box sizes of the present DNS: 
          flow field with puff-like structure, {\Large$\circ$}; 
          without puff, --.
          Bulk mean velocity, $u_{\rm m}$;
          centerline mean velocity, $u_{\rm c}$;
          box length in $i$-direction, $L_i$.
         }
 \vspace{-0.5em}
 \label{tab:condition}
 \begin{center}
 \small
  \begin{tabular}{l|cccccccccc}\hline
$Re_\tau$ ($=u_\tau \delta / \nu$)& 180  & 150  & 110  & 80   & 70   & 64   & 60$^\dagger$ 
                                  & 80   & 64 & 60 \\
Box size (cf. \Tref{tab:box})     & MB & MB & MB & MB & LB & LB & LB & XL & XL & XL \\\hline
$Re_{\rm m}$ ($=u_{\rm m} 2 \delta / \nu$)& 5680 & 4620 & 3270 & 2290 & 2000 & 1850 & --- 
                                          & 2310 & 1770 & 1640 \\
$Re_{\rm c}$ ($=u_{\rm c} \delta / \nu$)  & 3320 & 2710 & 1940 & 1400 & 1260 & 1200 & --- 
                                          & 1430 & 1140 & 1070 \\
with/without Puff & --- & --- & --- & --- & --- & --- & --- 
                  & \opencircle & \opencircle & \opencircle \\\hline
  \end{tabular}
 \normalsize
 \end{center}
 \vspace{-0.8em}
 \hspace{1em} \footnotesize{$\dagger$ This case resulted in a laminarization.\\}
\end{table}

\begin{figure}[tb]
 \begin{center}
  \epsfxsize=120.0mm
  \epsfbox{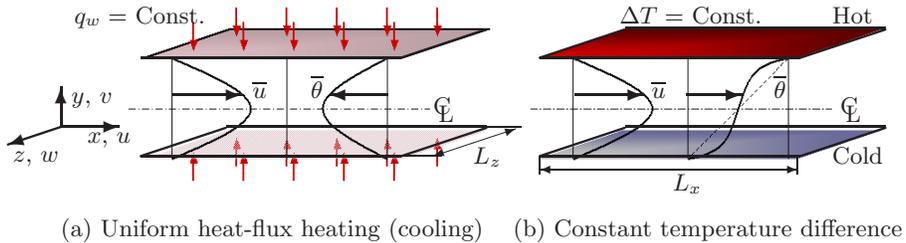}
 \end{center}
 \vspace{-1em}
 \caption{Configurations of the thermal boundary conditions.}
 \label{fig:domain}
\end{figure}

In this paper, the various statistics associated with fully-developed scalar fields for two different thermal boundary conditions are presented and discussed with emphasis on the role of the puff-like structure in the scalar transport.
A series of DNS has been made for $Re_\tau=60$--$180$ with a large computational box size, as summarized in Tables \ref{tab:condition} and \ref{tab:box}.

\section{Numerical procedure}

\begin{table}[t]
 \caption{Computational box size; $L_i$, $N_i$ and 
          $\Delta i^*$ (non-dimensionalized by $\delta$) are a box length, 
          a grid number and a spatial resolution of $i$-direction, respectively.
          }
 \vspace{-0.5em}
 \label{tab:box}
 \begin{center}
 \begin{tabular}{|c|c|c|c|}\hline
Box size                  & MB          & LB           & XL           \\\hline
$L_x\times L_y \times L_z$& $12.8\delta\times 2\delta \times  6.4\delta$ 
                          & $25.6\delta\times 2\delta \times 12.8\delta$ 
                          & $51.2\delta\times 2\delta \times 22.5\delta$ \\\hline
$N_x\times N_y \times N_z$& $ 512 \times$ $(96$--$128)$ $\times 256$ 
                          & $ 512 \times  96 \times 256$ 
                          & $1024 \times  96 \times 512$ \\\hline
$\Delta x^*, \Delta z^*$  & $0.05$, $0.025$ & $0.10$, $0.05$ & $0.05$, $0.044$ \\\hline
$\Delta y_{\tiny{\textrm{min}}}^*$--$\Delta y_{\tiny{\textrm{max}}}^*$
                          & $0.0011$--$0.033$ & $0.0014$--$0.045$ & $0.0014$--$0.045$ \\\hline
  \end{tabular}
 \end{center}
\end{table}

The mean flow of incompressible fluid was assumed to be driven by a uniform pressure gradient and we considered also passive scalar fields (see \Fref{fig:domain}).
One of the thermal boundary conditions is the uniform heat-flux heating over the both surfaces (UHF), and the other is the constant temperature difference between the walls (CTD).
The periodic boundary condition was imposed in the horizontal directions and the non-slip condition is applied on the walls. 
For the air of Prandtl number $Pr=0.71$, all fluid properties were treated as constant.
The fundamental equations are the continuity and the Navier-Stokes equations: 
\begin{equation}
 \frac{\partial u_i}{\partial x_i}=0,
 \label{eq:continuity}
\end{equation}
\begin{equation}
 \frac{\partial u_i^+}{\partial t^*} + u_j^+ \frac{\partial u_i^+}{\partial x_j^*}
 = -\frac{\partial p^+}{\partial x_i^*}
 + \frac{1}{Re_\tau} \frac{\partial^2 u_i^+}{\partial x_j^{*2}}
 + \delta_{1i},
 \label{eq:NS}
\end{equation}
where $\delta_{1i}$ corresponds to the mean pressure gradient. 
The energy equations are written as 
\begin{equation}
 \frac{\partial \theta^+}{\partial t^*} + u_j^+ \frac{\partial \theta^+}{\partial x_j^*}
 = \frac{1}{Re_\tau Pr} \frac{\partial^2 \theta^+}{\partial x_j^{*2}} + \frac{u^+}{u^+_\textrm{\tiny m}} 
 \hspace{1.0em} \textrm{ for UHF},
 \label{eq:energy1}
\end{equation}
\begin{equation}
 \frac{\partial \Theta^+}{\partial t^*} + u_j^+ \frac{\partial \Theta^+}{\partial x_j^*}
 = \frac{1}{Re_\tau Pr} \frac{\partial^2 \Theta^+}{\partial x_j^{*2}} -v^+
 \hspace{0.7em} \textrm{ for CTD},
 \label{eq:energy2}
\end{equation}
where in CTD, $\Theta$ ($=\theta/\Delta T-y^*$) is the deviation from the linear profile caused by the turbulence effect, and quantities with the superscript of ${}^+$ indicate those normalized by the wall variables and the friction temperature.
The last terms on the right-hand side of \Erefs{eq:energy1} and (\ref{eq:energy2}) represent a production by streamwise mean temperature gradient, $-u^+\partial_x T^+$, and a production by wall-normal temperature gradient of the linear profile, respectively.

For the spatial discretization, the finite difference method was adopted. 
Further details of the numerical scheme can be found in Kawamura \etal \cite{Kawa00}.
Uniform grid mesh was used in the horizontal directions, and non-uniform mesh in the $y$ direction.
A coarser mesh ($N_y=96$) was adopted for the low Reynolds numbers not over $Re_\tau=80$. 
At $Re_\tau=80$, the wall-normal grid spacings were $\Delta y^+=0.22$--$3.59$, which corresponded to $0.13\eta$--$1.25\eta$ ($\eta$ is referred to as a local Kolmogorov scale). 
These grid spacings were finer than ones used by Abe \etal \cite{Abe01}, whose grid resolutions were approximately equal to $0.3\eta$--$1.6\eta$.

A fully developed flow field at a higher $Re$ was successively used as the initial condition for a one-step lower $Re$.
Note that various statistical data and visualized fields were obtained after the scalar fields reached statistical-steady state.
As for two-dimensional contours shown after, quasi-mean velocity and temperature statistics were temporally averaged for a time of $150\delta/u_\tau$ ($9600\nu/u_\tau^2$, 40 wash-out times).
Further integration for $160\delta/u_\tau$ was necessary to obtain stable and accurate values of $C_f$ and \Nus presented here.

\section{Results and Discussion}
\subsection{Instantaneous fields}

\begin{figure}[t]
 \begin{center}
  \epsfxsize=150mm
  \epsfbox{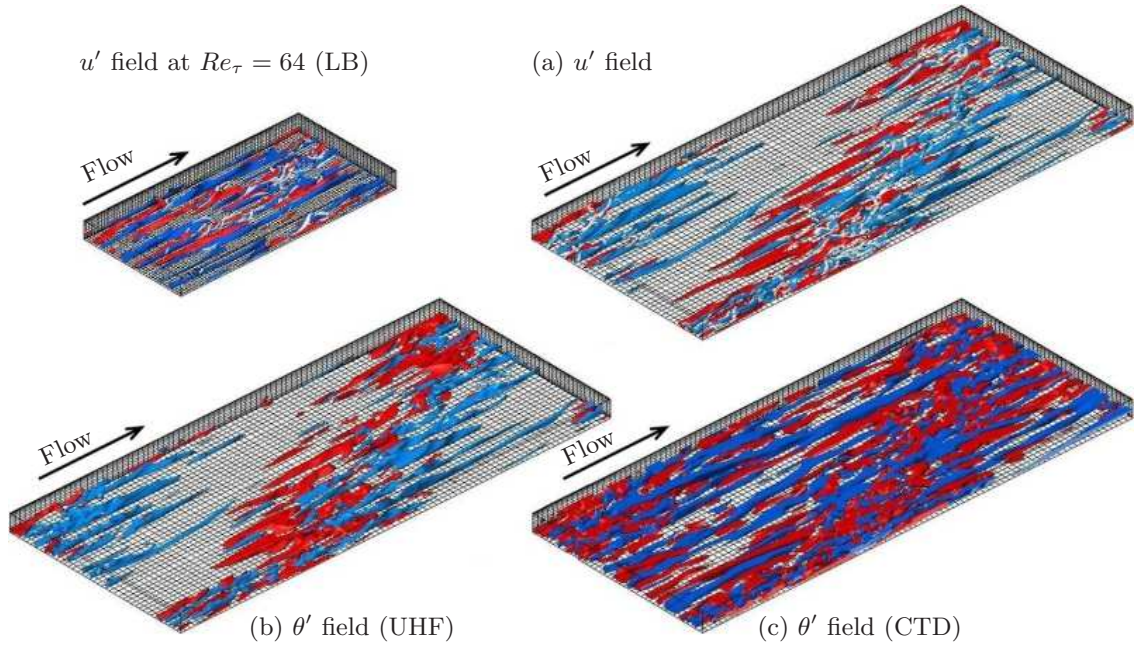}
 \caption{Instantaneous velocity and thermal fields for $Re_\tau=64$~(XL): high/low-speed ($a$) and high/low-temperature ($b$, $c$) regions (red, $u'^+$, $\theta'^+\ge3.0$; blue, $u'^+$, $\theta'^+\le-3.0$). White iso-surface in the flow field shows the contour of second invariant of the velocity gradient tensor: 
 $Q_2^+=\partial {u'_i}^+/\partial x_j^+ \cdot \partial {u'_j}^+/\partial x_i^+ \leq -0.03$. The visualized volume is the lower half of the computational box, namely $51.2\delta$$\times$$\delta$$\times$$22.5\delta$. A velocity field in the case of LB at the same Reynolds number is also shown for comparison.}
 \label{fig:instfld}
 \end{center}
\end{figure}

The puff-like structure as mentioned in the introduction was observed in the present DNS at even lower Reynolds numbers than $Re_\tau=80$.
\Fref{fig:instfld}($a$) shows the instantaneous flow field accompanied by the turbulent puff, which is sustained and equilibrium for $Re_\tau=60$--$80$~(XL).
The quasi-streamwise vortices are well represented by the negative $Q_2$ region, and the cluster of these fine-scale turbulent eddies is spatially isolated as seen in \Fref{fig:instfld}($a$).
Note that the streaks elongated in the streamwise direction penetrate the computational domain in the case of LB, indicating a shortage of the box size.
Moreover, fine-scale eddies are evenly distributed in the horizontal directions for LB.
Flatness factors of the velocity fluctuations are affected significantly by expanding the box size (not shown here).
Note also that in \Fref{fig:instfld}, the computed equilibrium puff stays constant in size, whereas a turbulent spot in a laminar channel flow is known to be widely spread with time \cite{Henningson87}.

Turning now to \Tref{tab:condition} obtained from the present DNS, it can be seen that a sustaining turbulence is obtained at as low as $Re_{\rm c}=1070$ ($Re_{\rm m}=1640$), which is consistent with the experimental results \cite{Patel69,Carlson82}.
At this Reynolds number, turbulence is spatially intermittent, that is, coexistence of the turbulent puff-like structure and less turbulence in space.
With decreasing $Re$, the intermittent flow is first observed at $Re_{\rm m}=2300$ in the present DNS: on the other hand, one may find in the literature that the intermittent flow in experiments occurred in the range of $1380<Re_{\rm m}<1800$, cf. \cite{Patel69}.

The isolated highly-disordered turbulent region, in which streaks are densely crowded, is spatially distributed in both $x$ and $z$ directions, while the turbulent puff of a pipe is intermittent only in the streamwise direction \cite{Wygnanski73}.
We have confirmed that this oblique structure in the channel flow was able to be captured even if an initial velocity field was a random distribution.
Thus one may regard the inclination of the puff-like structure as essential for a transitional channel flow.
As can be seen from \Tref{tab:condition}, we requires a large-scale domain such as XL to capture the puff in the channel flow.
Neither MB nor LB is large enough for the equilibrium to become established.
It is interesting to note that the self-sustaining puff in XL remains dominant with decreasing $Re_\tau$ from 64 to 60, whereas the turbulent flow has become laminar with LB.

From the visualization of the thermal fields affected by the puff, it is observed that thermal streaks in UHF (\Fref{fig:instfld}($b$)) are not homogeneously distributed and similar to the velocity field.
This is attributed to similarity of the boundary conditions.
In the cases of $u'$ and $\theta'$ for UHF, the visualized iso-surfaces of negative fluctuation show the low-speed (low-temperature) streaks in the wall region with an average spacing of $100$ wall units.
They are packed together and most of them are located at downstream of the highly-disordered turbulent region.
In addition, this dense clustering of negative (also positive) fluctuations clearly exhibits a very large-scale pattern in the UHF field.

In the case of CTD, both high- and low-temperature regions tend to exist uniformly in the core region (see \Fref{fig:instfld}($c$)).
Unlike UHF, the production of temperature variance for CTD is non-zero in the channel central region, and the location of maximum temperature variance is at the channel center.
The thermal structure in the core region is dominant rather than the near-wall streak as seen from \Fref{fig:instfld}($c$).
Thus, neither large-scale pattern nor the clustering of negative fluctuations is clearly observed in CTD with respect to puff-like structure.
This will be discussed later.

\subsection{Puff-like structure}
\subsubsection{Flow field}

\begin{figure}[p]
 \begin{center}
  \vspace{+0.7em} \hspace{+2.0em}
  \epsfxsize=145.0mm
  \epsfbox{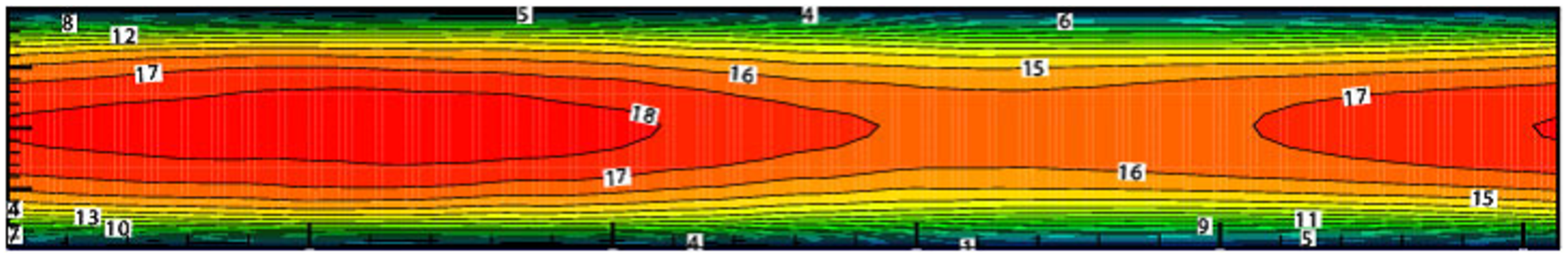}\\
  \vspace{-5.2em} \hspace{-40.0em} ($a$) \\
  \vspace{-2.5em} \hspace{-37.5em} $2\delta$ \\
  \vspace{+1.4em} \hspace{-37.0em} $\delta$ \\
  \vspace{+1.4em} \hspace{-37.0em} $0$ \\
  \vspace{+0.7em} \hspace{+2.0em}
  \epsfxsize=145.0mm
  \epsfbox{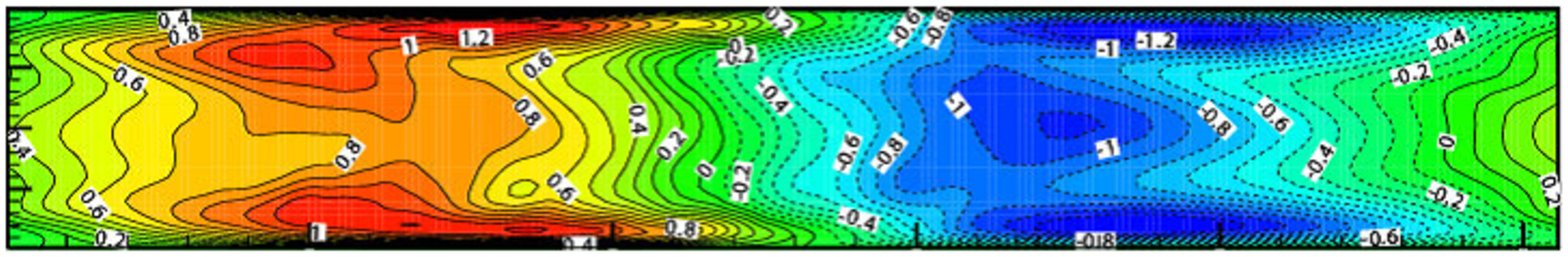}\\
  \vspace{-5.2em} \hspace{-40.0em} ($b$) \\
  \vspace{-2.5em} \hspace{-37.5em} $2\delta$ \\
  \vspace{+1.4em} \hspace{-37.0em} $\delta$ \\
  \vspace{+1.4em} \hspace{-37.0em} $0$ \\
  \vspace{+0.7em} \hspace{+2.0em}
  \epsfxsize=145.0mm
  \epsfbox{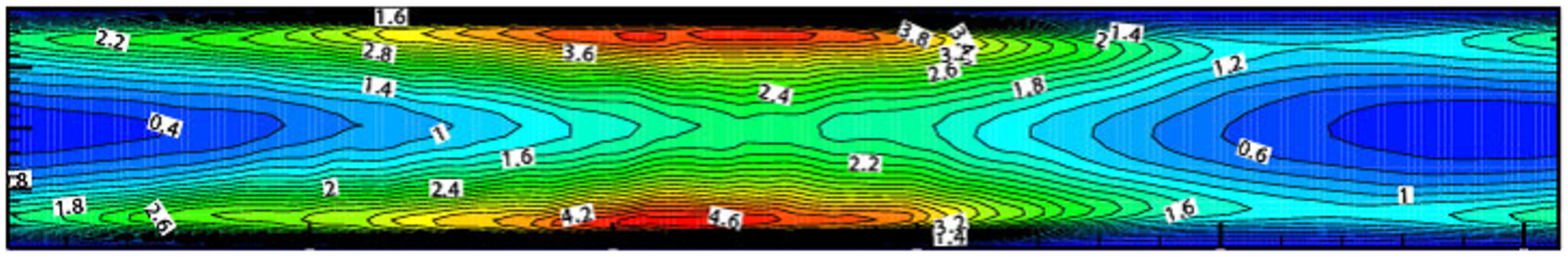}\\
  \vspace{-5.2em} \hspace{-40.0em} ($c$) \\
  \vspace{-2.5em} \hspace{-37.5em} $2\delta$ \\
  \vspace{+1.4em} \hspace{-37.0em} $\delta$ \\
  \vspace{+1.4em} \hspace{-37.0em} $0$ \\
  \vspace{+0.7em} \hspace{+2.0em}
  \epsfxsize=145.0mm
  \epsfbox{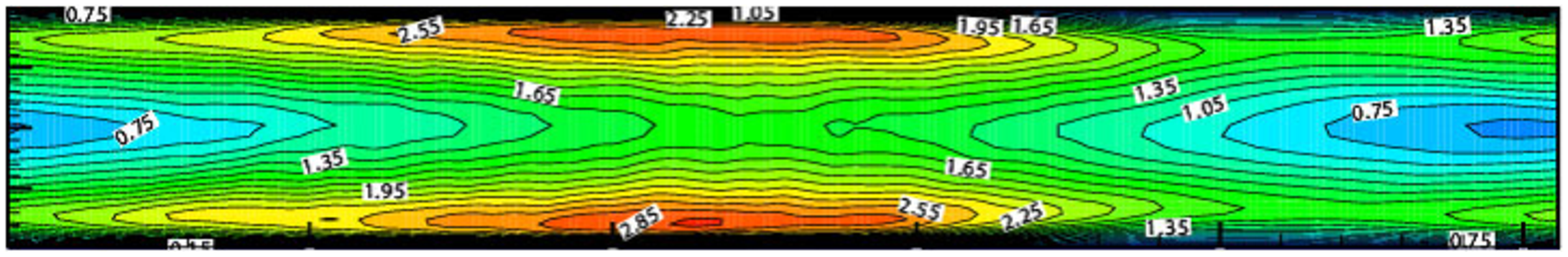}\\
  \vspace{-5.2em} \hspace{-40.0em} ($d$) \\
  \vspace{-2.5em} \hspace{-37.5em} $2\delta$ \\
  \vspace{+1.4em} \hspace{-37.0em} $\delta$ \\
  \vspace{+1.4em} \hspace{-37.0em} $0$ \\
  \vspace{+0.7em} \hspace{+2.0em}
  \epsfxsize=145.0mm
  \epsfbox{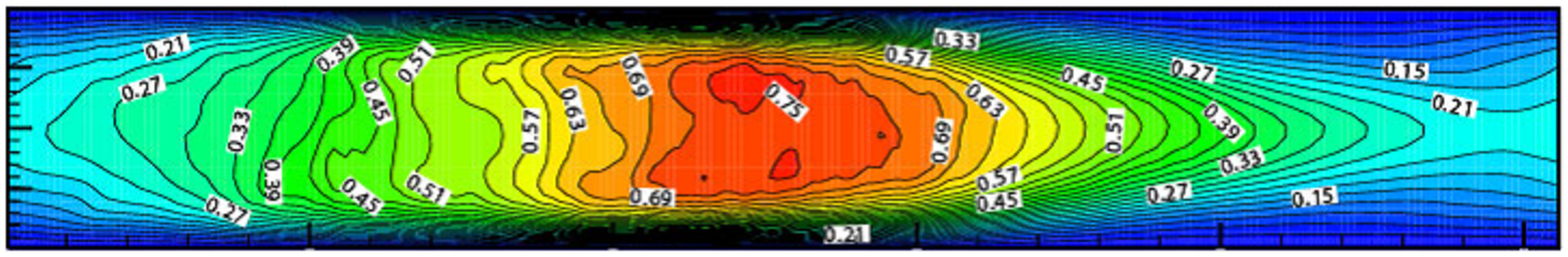}\\
  \vspace{-5.2em} \hspace{-40.0em} ($e$) \\
  \vspace{-2.5em} \hspace{-37.5em} $2\delta$ \\
  \vspace{+1.4em} \hspace{-37.0em} $\delta$ \\
  \vspace{+1.4em} \hspace{-37.0em} $0$ \\
  \vspace{+0.7em} \hspace{+2.0em}
  \epsfxsize=145.0mm
  \epsfbox{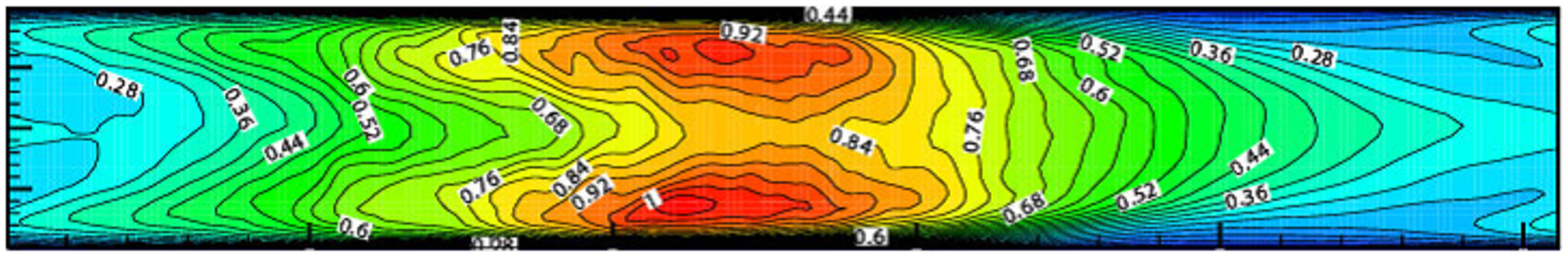}\\
  \vspace{-5.2em} \hspace{-40.0em} ($f$) \\
  \vspace{-2.5em} \hspace{-37.5em} $2\delta$ \\
  \vspace{+1.4em} \hspace{-37.0em} $\delta$ \\
  \vspace{+1.4em} \hspace{-37.0em} $0$ \\
  \vspace{+0.7em} \hspace{+2.0em}
  \epsfxsize=145.0mm
  \epsfbox{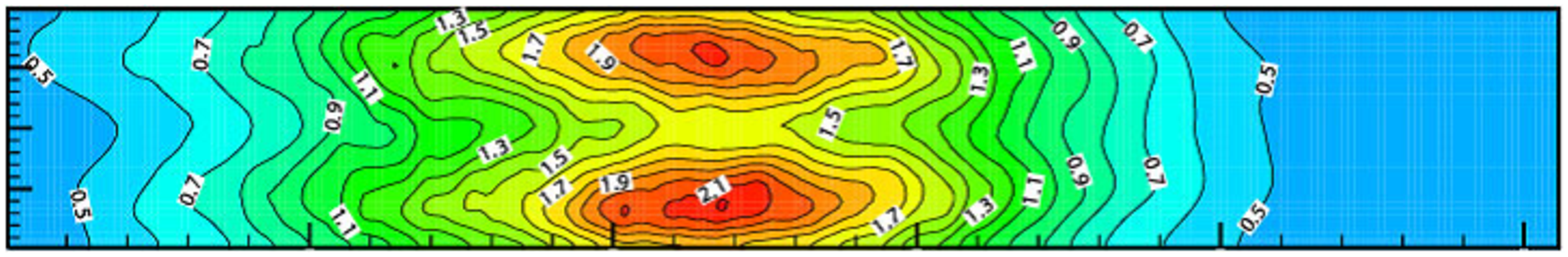}\\
  \vspace{-5.2em} \hspace{-40.0em} ($g$) \\
  \vspace{-2.5em} \hspace{-37.5em} $2\delta$ \\
  \vspace{+1.4em} \hspace{-37.0em} $\delta$ \\
  \vspace{+1.4em} \hspace{-37.0em} $0$ \\
  \begin{picture}(420,1)
   \put(122.4,4){\line(0,1){530}}
   \put(199.4,4){\line(0,1){530}}
   \put(299.0,4){\line(0,1){530}}
   \put(421.0,4){\line(0,1){530}}
   \put( 18,11){\vector(1,0){104.4}} \put(430,11){\vector(-1,0){9}}
   \put(150,11){\vector(1,0){49.4}} \put(150,11){\vector(-1,0){27.6}}
   \put(250,11){\vector(1,0){49.0}} \put(250,11){\vector(-1,0){51.5}}
   \put(350,11){\vector(1,0){71.0}} \put(350,11){\vector(-1,0){51}}
   \thicklines
   \put(0,-2){\vector(1,0){40}} \put(0,-2){\vector(0,1){25}}
  \end{picture}\\
  \vspace{-3.7em} \hspace{-39em} $y$ \\
  \vspace{+2.0em} \hspace{-30em} $x$ \\
  {\small \vspace{-2.1em} 
     \hspace{-0em} $\Delta x=14\delta$ 
     \hspace{4.5em} $9.6\delta$ 
     \hspace{5.5em} $12.4\delta$ 
     \hspace{7.2em} $15.2\delta$} \\
  {\small \vspace{-0.8em} 
     \hspace{+11.3em} I 
     \hspace{6.1em} \II
     \hspace{8.2em} \III 
     \hspace{10.0em} \IV} \\
  \vspace{0.5em}
  \Fref{fig:puff2d_u} For legend see page \pageref{fig:puff2d_u}.
 \end{center}
\end{figure}

\begin{figure}[t]
 \begin{center}
  \vspace{+0.7em} \hspace{+2.0em}
  \epsfxsize=145.0mm
  \epsfbox{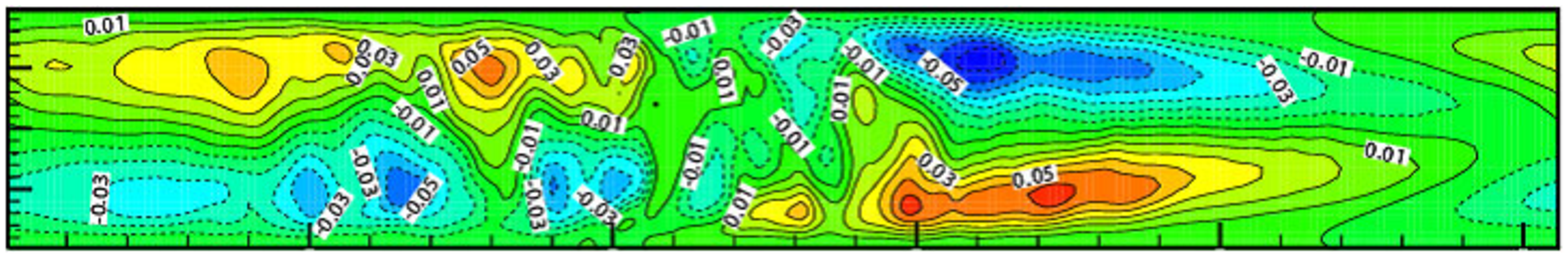}\\
  \vspace{-5.0em} \hspace{-40.0em} ($h$) \\
  \vspace{-2.5em} \hspace{-37.5em} $2\delta$ \\
  \vspace{+1.4em} \hspace{-37.0em} $\delta$ \\
  \vspace{+1.4em} \hspace{-37.0em} $0$ \\
  \vspace{+0.7em} \hspace{+2.0em}
  \epsfxsize=145.0mm
  \epsfbox{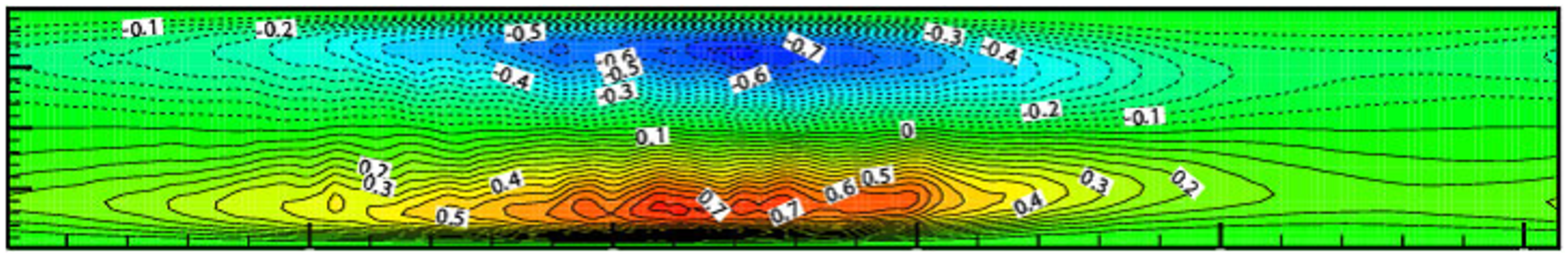}\\
  \vspace{-5.0em} \hspace{-40.0em} ($i$) \\
  \vspace{-2.5em} \hspace{-37.5em} $2\delta$ \\
  \vspace{+1.4em} \hspace{-37.0em} $\delta$ \\
  \vspace{+1.4em} \hspace{-37.0em} $0$ \\
  \vspace{+0.7em} \hspace{+2.0em}
  \epsfxsize=145.0mm
  \epsfbox{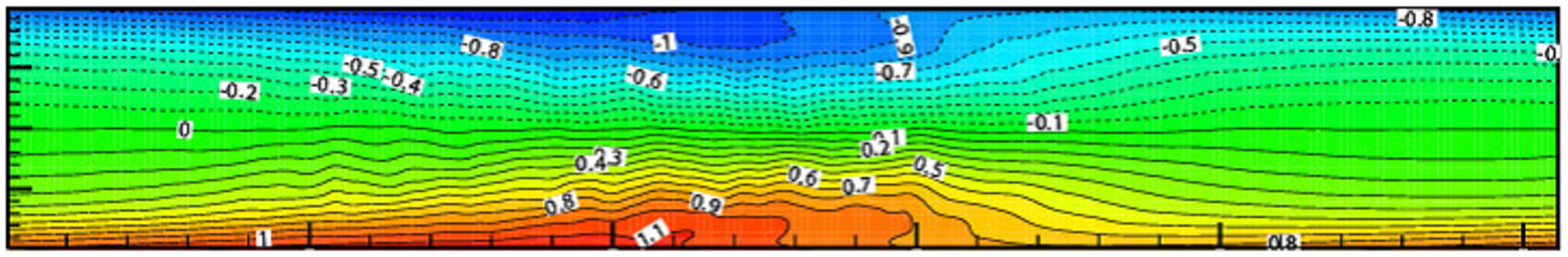}\\
  \vspace{-5.0em} \hspace{-40.0em} ($j$) \\
  \vspace{-2.5em} \hspace{-37.5em} $2\delta$ \\
  \vspace{+1.4em} \hspace{-37.0em} $\delta$ \\
  \vspace{+1.4em} \hspace{-37.0em} $0$ \\
  \begin{picture}(420,1)
   \put(122.4,4){\line(0,1){234}}
   \put(199.4,4){\line(0,1){234}}
   \put(299.0,4){\line(0,1){234}}
   \put(421.0,4){\line(0,1){234}}
   \put( 18,11){\vector(1,0){104.4}} \put(430,11){\vector(-1,0){9}}
   \put(150,11){\vector(1,0){49.4}} \put(150,11){\vector(-1,0){27.6}}
   \put(250,11){\vector(1,0){49.0}} \put(250,11){\vector(-1,0){51.5}}
   \put(350,11){\vector(1,0){71.0}} \put(350,11){\vector(-1,0){51}}
   \thicklines
   \put(0,-2){\vector(1,0){40}} \put(0,-2){\vector(0,1){25}}
  \end{picture}\\
  \vspace{-3.7em} \hspace{-39em} $y$ \\
  \vspace{+2.0em} \hspace{-30em} $x$ \\
  {\small \vspace{-2.1em} 
     \hspace{-0em} $\Delta x=14\delta$ 
     \hspace{4.5em} $9.6\delta$ 
     \hspace{5.5em} $12.4\delta$ 
     \hspace{7.2em} $15.2\delta$} \\
  {\small \vspace{-0.8em} 
     \hspace{+11.3em} I 
     \hspace{6.1em} \II
     \hspace{8.2em} \III 
     \hspace{10.0em} \IV} \\
  \caption{Quasi-mean flow field 
           in an ($x$-$y$) frame of reference moving with the puff at $Re_\tau=64$ (XL): 
           the ensemble-averaged pattern of 
           ($a$) quasi-mean streamwise velocity $\overline{u}^{z'}$ , 
           ($b$) $\overline{u}^{z'} - \overline{u}$, 
           ($c$) turbulent kinetic energy $k''=\overline{u''_iu''_i/2}^{z'}$, 
           ($d$) turbulence intensity $u''_{\rm rms}$, 
           ($e$) $v''_{\rm rms}$, 
           ($f$) $w''_{\rm rms}$, 
           ($g$) $p''_{\rm rms}$, 
           ($h$) wall-normal component of quasi-mean velocity $\overline{v}^{z'}$ , 
           ($i$) Reynolds shear stress $-\overline{u''v''}^{z'}$, 
           and ($j$) shear stress $\overline{\tau}^{z'}$ of \Eref{eq:shear}, 
           where all of quantities are normalized by $u_\tau$.
           Mean-flow direction is from left to right.
           Solid and dashed lines represent positive and negative quantities, respectively.
           The vertical lines are located at the streamwise locations where 
           $\overline{u_{\rm c}}^{z'}$ is maximum (I) or minimum (\III); 
           $k''$ maximum (\II) or minimum(\IV).
          }
  \label{fig:puff2d_u}
 \end{center}
 \vspace{-0.67em}
\end{figure}

\begin{figure}[t]
 \begin{center}
  \vspace{+0.7em} \hspace{+2.0em}
  \epsfxsize=145.0mm
  \epsfbox{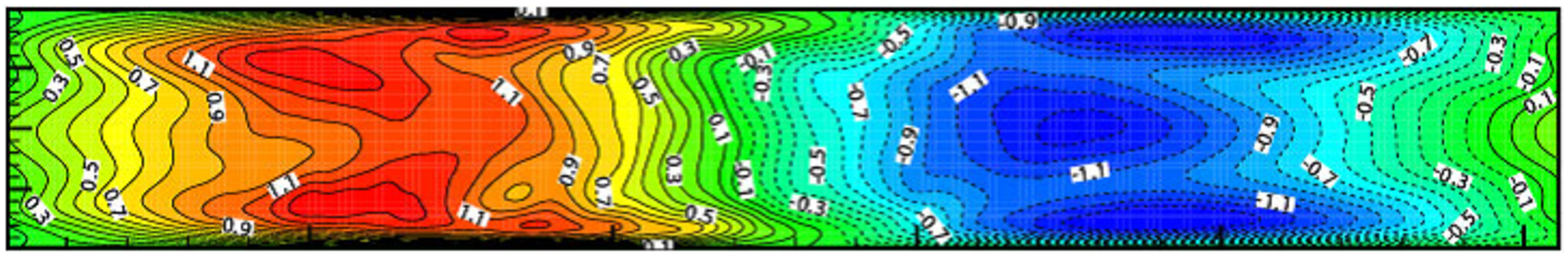}\\
  \vspace{-5.0em} \hspace{-40.0em} ($a$) \\
  \vspace{-2.5em} \hspace{-37.5em} $2\delta$ \\
  \vspace{+1.4em} \hspace{-37.0em} $\delta$ \\
  \vspace{+1.4em} \hspace{-37.0em} $0$ \\
  \vspace{+0.7em} \hspace{+2.0em}
  \epsfxsize=145.0mm
  \epsfbox{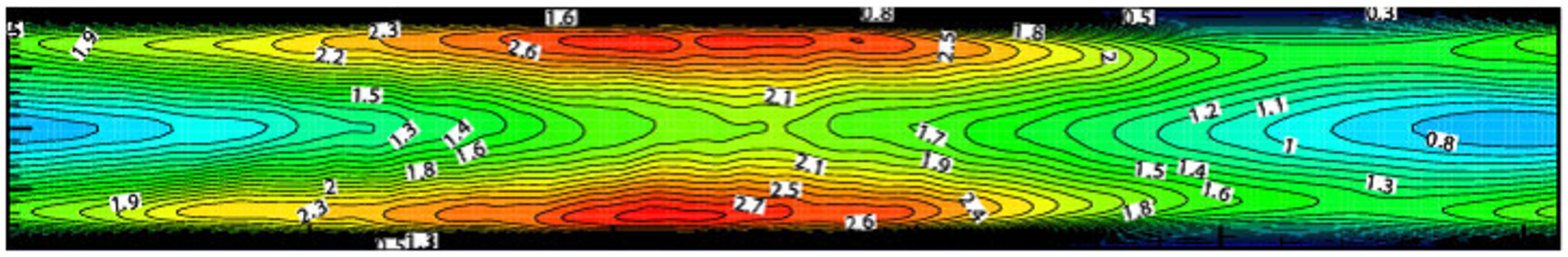}\\
  \vspace{-5.0em} \hspace{-40.0em} ($b$) \\
  \vspace{-2.5em} \hspace{-37.5em} $2\delta$ \\
  \vspace{+1.4em} \hspace{-37.0em} $\delta$ \\
  \vspace{+1.4em} \hspace{-37.0em} $0$ \\
  \vspace{+0.7em} \hspace{+2.0em}
  \epsfxsize=145.0mm
  \epsfbox{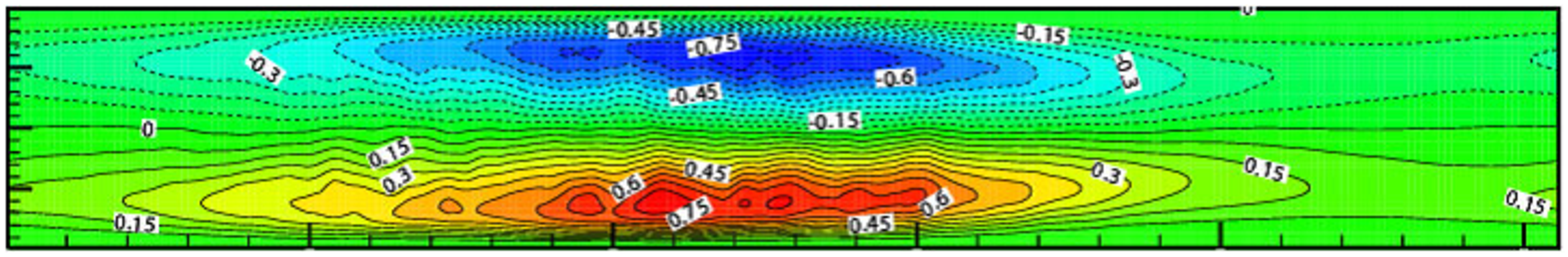}\\
  \vspace{-5.0em} \hspace{-40.0em} ($c$) \\
  \vspace{-2.5em} \hspace{-37.5em} $2\delta$ \\
  \vspace{+1.4em} \hspace{-37.0em} $\delta$ \\
  \vspace{+1.4em} \hspace{-37.0em} $0$ \\
  \begin{picture}(420,1)
   \put(122.4,4){\line(0,1){234}}
   \put(199.4,4){\line(0,1){234}}
   \put(299.0,4){\line(0,1){234}}
   \put(421.0,4){\line(0,1){234}}
   \put( 18,11){\vector(1,0){104.4}} \put(430,11){\vector(-1,0){9}}
   \put(150,11){\vector(1,0){49.4}} \put(150,11){\vector(-1,0){27.6}}
   \put(250,11){\vector(1,0){49.0}} \put(250,11){\vector(-1,0){51.5}}
   \put(350,11){\vector(1,0){71.0}} \put(350,11){\vector(-1,0){51}}
   \thicklines
   \put(0,-2){\vector(1,0){40}} \put(0,-2){\vector(0,1){25}}
  \end{picture}\\
  \vspace{-3.7em} \hspace{-39em} $y$ \\
  \vspace{+2.0em} \hspace{-30em} $x$ \\
  {\small \vspace{-2.1em} 
     \hspace{-0em} $\Delta x=14\delta$ 
     \hspace{4.5em} $9.6\delta$ 
     \hspace{5.5em} $12.4\delta$ 
     \hspace{7.2em} $15.2\delta$} \\
  {\small \vspace{-0.8em} 
     \hspace{+11.3em} I 
     \hspace{6.1em} \II
     \hspace{8.2em} \III 
     \hspace{10.0em} \IV} \\
  \caption{Quasi-mean thermal field of UHF
           in an ($x$-$y$) frame of reference moving with the puff at $Re_\tau=64$ (XL): 
           the ensemble-averaged pattern of 
           ($a$) deviation between \emph{quasi}-mean and mean temperature 
            $(\overline{\theta}^{z'} - \overline{\theta})$, 
           ($b$) temperature variance $\theta''_{\rm rms}$, 
           and ($c$) wall-normal turbulent heat flux $-\overline{v''\theta''}^{z'}$, 
           where all of quantities are normalized with a friction temperature
           (and a friction velocity).
           Mean-flow direction is from left to right.
           Solid and dashed lines represent positive and negative quantities, respectively.
           Positions of the vertical lines with the numerals I--\IVs 
           are same as those in \Fref{fig:puff2d_u}.
          }
  \label{fig:puff2d_t1}
 \end{center}
\end{figure}

\begin{figure}[t]
 \begin{center}
  \vspace{+0.7em} \hspace{+2.0em}
  \epsfxsize=145.0mm
  \epsfbox{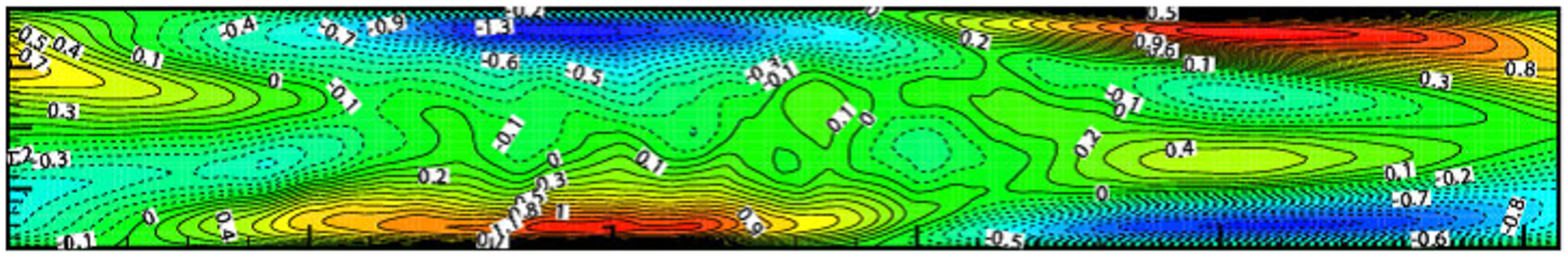}\\
  \vspace{-5.0em} \hspace{-40.0em} ($a$) \\
  \vspace{-2.5em} \hspace{-37.5em} $2\delta$ \\
  \vspace{+1.4em} \hspace{-37.0em} $\delta$ \\
  \vspace{+1.4em} \hspace{-37.0em} $0$ \\
  \vspace{+0.7em} \hspace{+2.0em}
  \epsfxsize=145.0mm
  \epsfbox{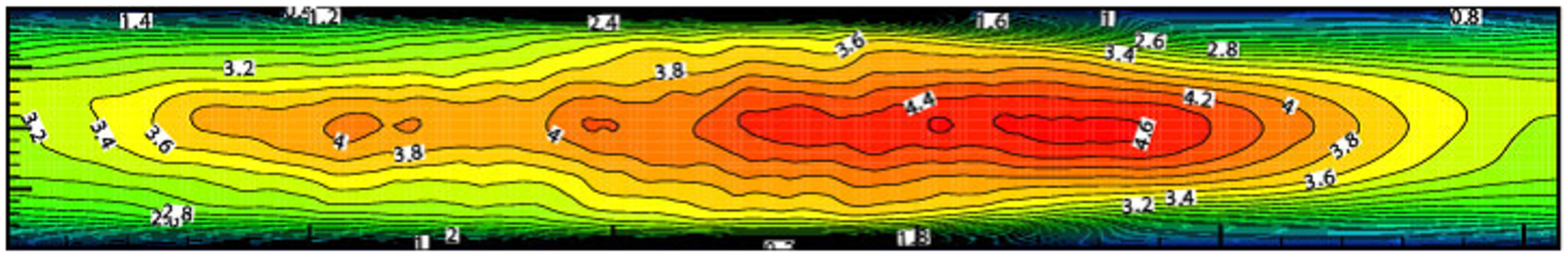}\\
  \vspace{-5.0em} \hspace{-40.0em} ($b$) \\
  \vspace{-2.5em} \hspace{-37.5em} $2\delta$ \\
  \vspace{+1.4em} \hspace{-37.0em} $\delta$ \\
  \vspace{+1.4em} \hspace{-37.0em} $0$ \\
  \vspace{+0.7em} \hspace{+2.0em}
  \epsfxsize=145.0mm
  \epsfbox{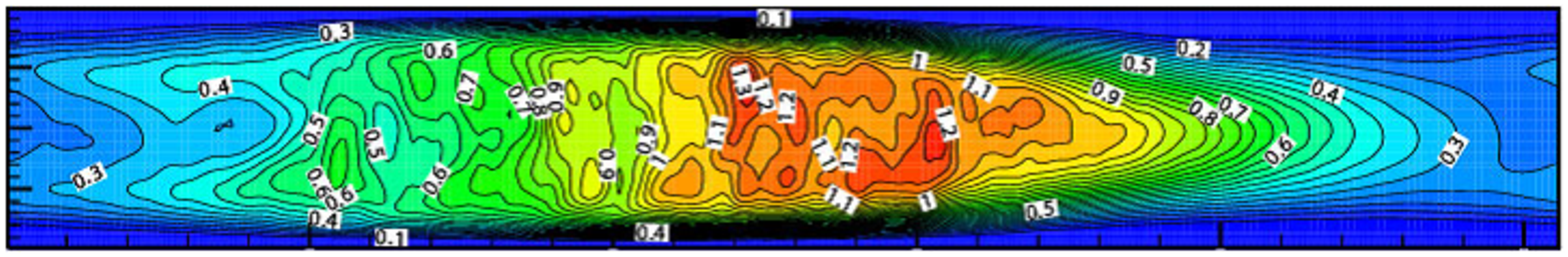}\\
  \vspace{-5.0em} \hspace{-40.0em} ($c$) \\
  \vspace{-2.5em} \hspace{-37.5em} $2\delta$ \\
  \vspace{+1.4em} \hspace{-37.0em} $\delta$ \\
  \vspace{+1.4em} \hspace{-37.0em} $0$ \\
  \begin{picture}(420,1)
   \put(122.4,4){\line(0,1){234}}
   \put(199.4,4){\line(0,1){234}}
   \put(299.0,4){\line(0,1){234}}
   \put(421.0,4){\line(0,1){234}}
   \put( 18,11){\vector(1,0){104.4}} \put(430,11){\vector(-1,0){9}}
   \put(150,11){\vector(1,0){49.4}} \put(150,11){\vector(-1,0){27.6}}
   \put(250,11){\vector(1,0){49.0}} \put(250,11){\vector(-1,0){51.5}}
   \put(350,11){\vector(1,0){71.0}} \put(350,11){\vector(-1,0){51}}
   \thicklines
   \put(0,-2){\vector(1,0){40}} \put(0,-2){\vector(0,1){25}}
  \end{picture}\\
  \vspace{-3.7em} \hspace{-39em} $y$ \\
  \vspace{+2.0em} \hspace{-30em} $x$ \\
  {\small \vspace{-2.1em} 
     \hspace{-0em} $\Delta x=14\delta$ 
     \hspace{4.5em} $9.6\delta$ 
     \hspace{5.5em} $12.4\delta$ 
     \hspace{7.2em} $15.2\delta$} \\
  {\small \vspace{-0.8em} 
     \hspace{+11.3em} I 
     \hspace{6.1em} \II
     \hspace{8.2em} \III 
     \hspace{10.0em} \IV} \\
  \caption{Same as \Fref{fig:puff2d_t1} but for CTD.
          }
  \label{fig:puff2d_t2}
 \end{center}
\end{figure}

The puff-like structure emerges spontaneously from featureless turbulence: however, its oblique band is confined by the periodic boundary in both $x$ and $z$ directions, and the band parallels with a diagonal of the domain, as given in \Fref{fig:instfld}.
Accordingly, let us define $z'$ as a coordinate that is parallel to the diagonal line.
A value spatially-averaged in the $z'$ direction and a fluctuation from this \emph{quasi}-mean value are defined as: 
\begin{equation}
 \overline{u_i}^{z'} (x,y) = \frac{1}{T}\frac{1}{L_{z'}}
 \int_0^{T} \!\!\!\int_0^{L_{z'}} u_i(x+ u_{\rm m}t,y,z',t){\rm d}z'{\rm d}t,
 \hspace{1em} u''_i=u_i - \overline{u_i}^{z'},
\end{equation}
\begin{equation}
 \overline{\theta}^{z'} (x,y) = \frac{1}{T}\frac{1}{L_{z'}}
 \int_0^{T} \!\!\!\int_0^{L_{z'}} \theta (x+ u_{\rm m}t,y,z',t){\rm d}z'{\rm d}t,
 \hspace{1em} \theta''=\theta - \overline{\theta}^{z'}. 
 \label{eq:tave}
\end{equation}
Here, the propagation velocity of the puff was found, in the previous work \cite{Tsukahara05}, to be constant and almost equal to the bulk mean velocity $u_{\rm m}$. 
By assuming homogeneity of the puff in the $z'$ direction, i.e. the direction from $(x,z)=(0,L_z)$ to $(L_x,0)$ in \Fref{fig:instfld}, the ensemble-averaged velocity and temperature fields with respect to the puff are obtained at $Re_\tau=64$~(XL) as given in \Frefs{fig:puff2d_u}--\ref{fig:puff2d_t2}. 
In these figures, one should keep in mind that the vertical scale is four times the abscissa scale.
An overline of $\overline{(\cdot)}$ denotes the spatial (in both $x$ and $z$ directions) and temporal averaging.

\Fref{fig:puff2d_u}($b$) shows that the large-scale regions of high- and low-speed fluctuations emerge occupying the whole width in the wall-normal direction.
At the channel centerline, the quasi-mean velocity $\overline{u_{\rm c}}^{z'}$ becomes maximum (on the line I in \Fref{fig:puff2d_u}) about $22\delta$ upstream from the position of minimum $\overline{u_{\rm c}}^{z'}$ (\III).
This streamwise scale of the puff-like structure is almost same as that of an 
equilibrium puff in a transitional pipe (see, e.g., \cite{Priymak04}).
Between the upstream high-momentum (around the vertical-line I) and the downstream low-momentum regions (\III), an intensive-turbulence region (\II) appears as seen in \Fref{fig:puff2d_u}($c$).
On the other hand, around the line \IVs (called a `region \IV', hereafter), all the components of the turbulence intensity and the pressure fluctuation are attenuated as given in \Frefs{fig:puff2d_u}($d$--$g$). 
Since the intensity of $u''$ is remarkably larger than those of the other two components, the contour of $k''$ (\Fref{fig:puff2d_u}($c$)) resemble that of $u''$.
The spatial extent of the reductions in $k''$ and $u''$ is coincided with the local acceleration of the mean flow in the region \IV.

The quasi-mean $v$-velocity is as large as $\pm0.05$ just around the turbulent region (see \Fref{fig:puff2d_u}($h$)), whereas in the fully turbulent channel flow the mean $v$ must be zero.
The high-momentum fluid (around I of \Fref{fig:puff2d_u}($b$)) impinges on the downstream low-momentum one and is swept towards the walls in the upstream extent of the interface, $\sim$I$\sim$\II. 
On the contrary, one can observe that the fluid converges towards the channel center as it crosses the interface \II$\sim$\III$\sim$. 

The contours of the Reynolds shear stress and total one (defined by the following equation) are shown in \Frefs{fig:puff2d_u}($i$) and ($j$), respectively.
\begin{equation}
 \overline{\tau}^{z'}=-\overline{u}^{z'}\overline{v}^{z'}-\overline{u''v''}^{z'}
 +\nu \frac{\partial \overline{u}^{z'}}{\partial y}.
 \label{eq:shear}
\end{equation}
Here, $-\overline{u}^{z'}\overline{v}^{z'}$ is very small compared to $-\overline{u''v''}^{z'}$, so that $-\overline{u}^{z'}\overline{v}^{z'}$ is neglected here.
A fully turbulent channel flow is homogeneous in the horizontal directions, so that the all derivatives with respect to $x$ and $z$ can be assumed to be zero, except for the pressure gradient $\partial_x \overline{p}$, which drives the mean flow, cf. \cite{Tennekes72}.
On the other hand, the flow with the puff is homogeneous only in the $z'$ direction, so that the total shear stress is not strictly defined as \Eref{eq:shear}.
However, all derivatives with respect to $x$ and $z$ (not shown here) are small enough compared to the terms of \Eref{eq:shear}. 
The maximum Reynolds stress is located at about $y=0.4\delta$ in the region \IIs, where the secondary flow appears with significant non-zero $\overline{v}^{z'}$.
Even though $\overline{v}^{z'}$ is small, its effect on the velocity $\overline{u}^{z'}$ at the near-wall region is large.
In consequence, the wall shear stress (and the skin friction coefficient) changes about 40\% 
in the streamwise direction.

\subsubsection{Thermal fields}

Using \Eref{eq:tave}, the averaged properties for the thermal fields of UHF and CTD are given in \Frefs{fig:puff2d_t1} and \ref{fig:puff2d_t2}, respectively.
\Fref{fig:puff2d_t1}($a$) shows the contour of the quasi-mean temperature for UHF, indicating the existence of large-scale structure (high- and low-$\overline{\theta}^{z'}$ regions) similar to that of the velocity field shown in \Frefs{fig:puff2d_u}($b$). 
On the other hand, such structure cannot be found in the case of CTD (\Fref{fig:puff2d_t2}($a$)).
Also shown in \Frefs{fig:puff2d_t1}($b$) and \ref{fig:puff2d_t2}($b$) are the root-mean-square fluctuations $\theta''_{\rm rms}$ for each temperature fields.
The difference in the distributions of $\theta''_{\rm rms}$ between the two thermal boundary conditions results from the difference in each distribution of $\overline{v''\theta''}^{z'}$.

\Fref{fig:puff2d_t1}($c$) shows the contour of $\overline{v''\theta''}^{z'}$ for UHF, it displays intermittent regions of large $\overline{v''\theta''}^{z'}$ similar to the spatial distribution of $\overline{u''v''}^{z'}$.
This implies that strong production of the scalar fluctuation also takes place intermittently just as that of the velocity fluctuation.
As for CTD, the intermittent distribution of $\overline{v''\theta''}^{z'}$ is clearly identified but large $\overline{v''\theta''}^{z'}$ is located at the channel center and slightly downstream from the line \II, as given in \Fref{fig:puff2d_t2}($c$).
Both \Frefs{fig:puff2d_t1}($c$) and \ref{fig:puff2d_t2}($c$) indicate that around the line \II, the wall-normal heat flux becomes as large as twice of its mean values (not shown here) in both of the thermal boundary conditions.
This spatial intermittency of the effective heat transport is associated with the wall-normal secondary flow (see \Fref{fig:puff2d_u}($h$)).
The flow pattern of the secondary flow is symmetric about the channel centerline whereas the wall-normal mean temperature profile of CTD is asymmetric. Hence a streamwise variation of $\overline{\theta}^{z'}$ is negligible at the channel center as seen from \Fref{fig:puff2d_t2}($a$).
As a result, no large-scale structure of $\theta'$ for CTD is found in the visualization shown in \Fref{fig:instfld}($c$).
In the near-wall region, however, \Fref{fig:puff2d_t2}($a$) shows significant variation of $\overline{\theta}^{z'}$ in the $x$ direction, see also \Fref{fig:puff2d_t1}($a$) for UHF.
The wall heat flux $\overline{q_\textrm{\tiny wall}}^{z'} (=\partial_y \overline{\theta}^{z'})$ also changes about 60\%
in the $x$ direction for both thermal boundary conditions.
This reveals that heat transfer (and, therefore, Nusselt number) is spatially enhanced by the puff.


\begin{figure}[t]
 \vspace{+0.5em}
 \begin{tabular}{cc}
  \hspace{-0.7em}
  \begin{minipage}[t]{0.46\textwidth}
   \begin{center}
    \epsfysize=97.0mm
    \epsfbox{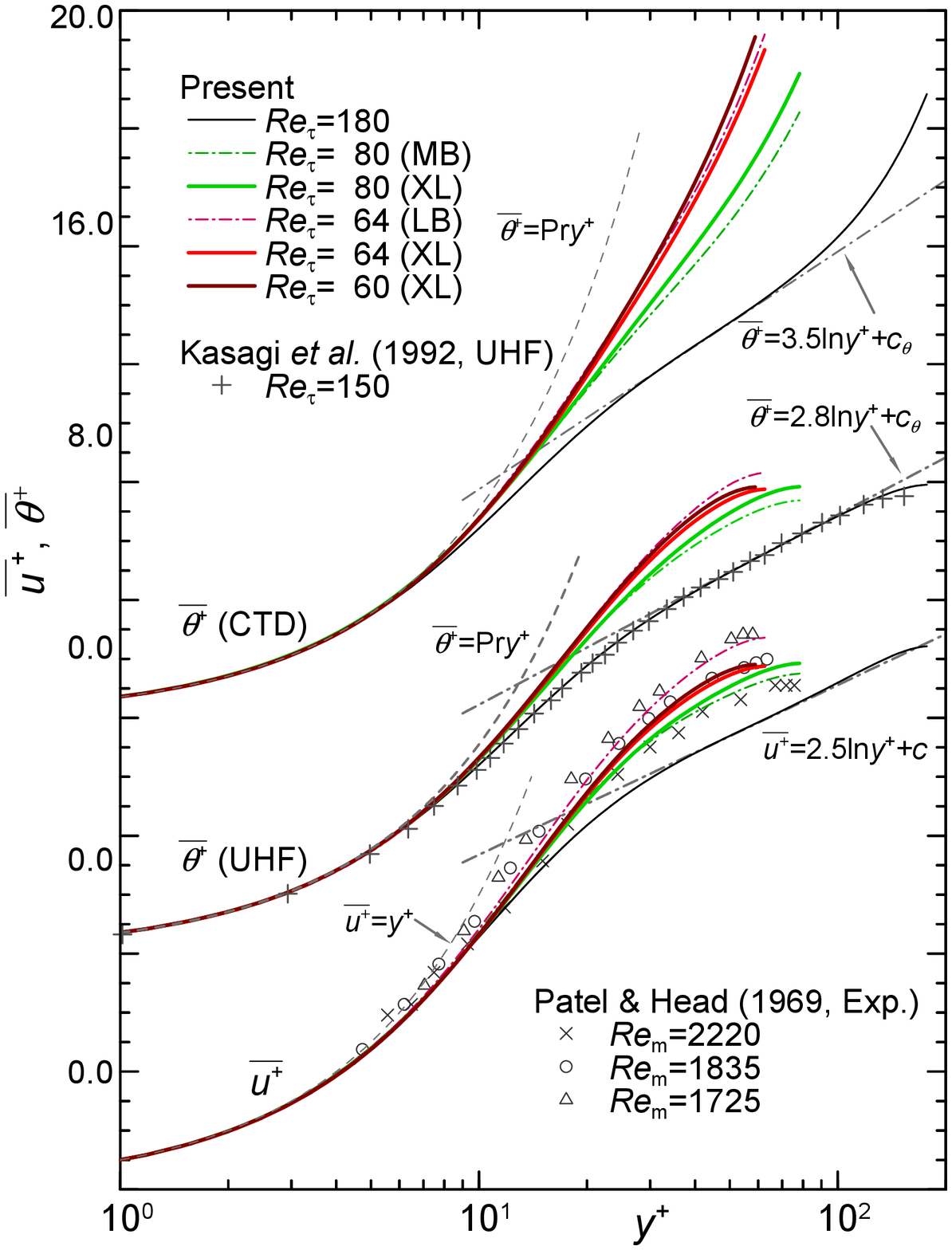}\\
    \vspace{-0.6em}
    \caption{Mean velocity $\overline{u}$ 
             and temperature $\overline{\theta}$ profiles 
             in viscous wall-units.
            }
    \label{fig:ut12_mean}
   \end{center}
  \end{minipage}
  \hspace{0.3em}
  \begin{minipage}[t]{0.51\textwidth}
   \begin{center}
   \hspace{-0.4em}
    \epsfysize=97.0mm
    \epsfbox{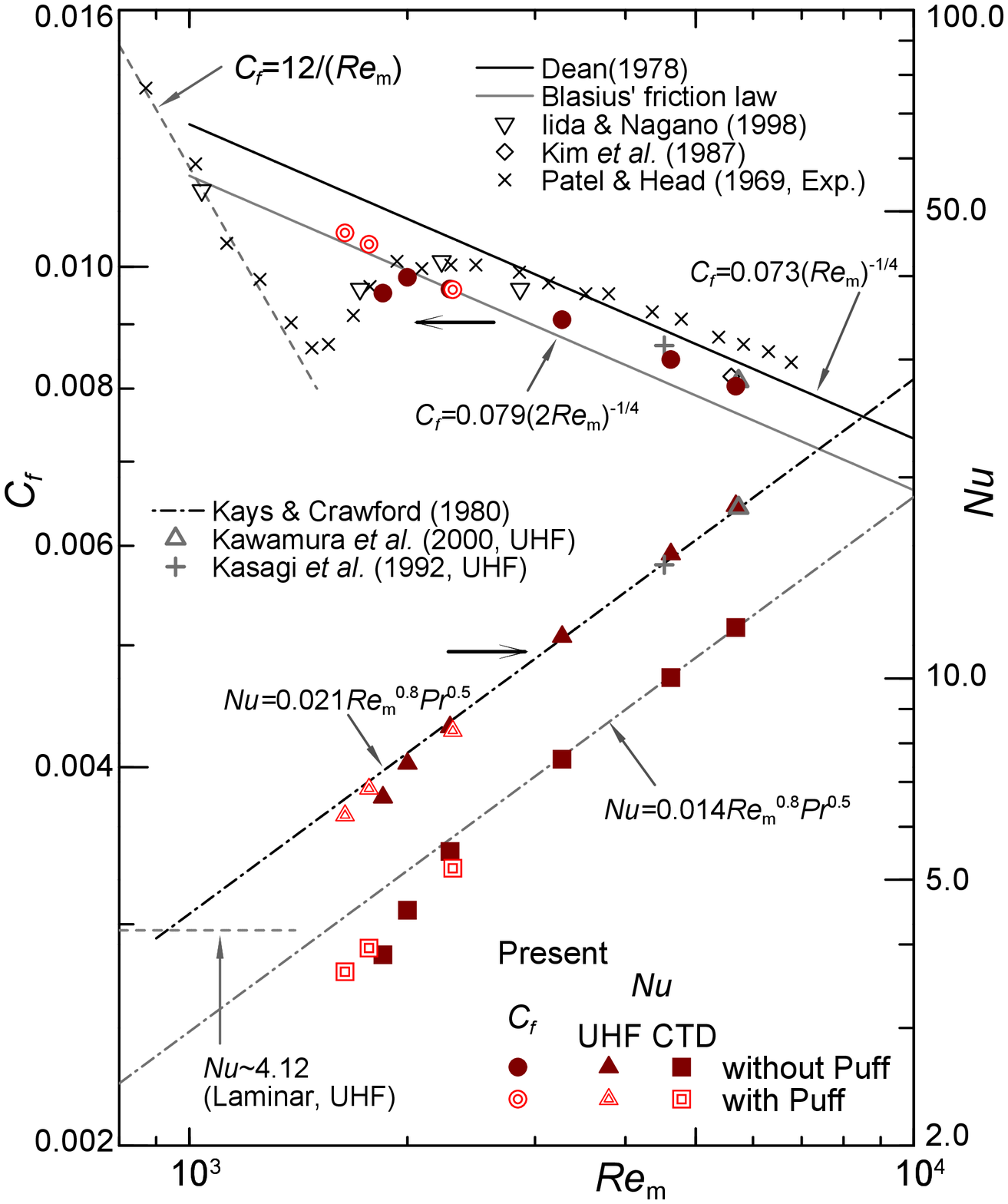}\\
    \vspace{-5.1em} $\overbrace{\hspace{2.8em}}^{\ }$ \hspace{-6em}\,\\
    \vspace{2.0em}
    \hspace{-0.2em}
    \caption{Variation with Reynolds number of $C_f$ and \Nu.
             Laminar flow relations, (-- -- --). 
            }
    \label{fig:cf_nu}
   \end{center}
  \end{minipage}
 \end{tabular}
\end{figure}

\subsection{Mean flow parameter}

The dimensionless mean velocity and temperature profiles are shown in \Fref{fig:ut12_mean}.
For $Re_\tau\leq80$, the Reynolds-number dependence of the mean velocity profile in the present DNS data is consistent with that by Patel \& Head \cite{Patel69}.
The friction Reynolds number of the experiment was set to be $Re_\tau=59$--$77$. 
If emphasis is placed on the data at $Re_\tau=80$~(MB and XL), the profiles suggest that the maximum (channel-centerline) values of $\overline{u}^+$ and $\overline{\theta}^+$ increase with extending the box size up to XL, since the quasi-laminar region locally appeared in the flow field.
However, the Reynolds-number dependencies of both mean velocity and temperature profiles in the flow involving the puff are significantly smaller than those without the puff.
For instance, the centerline velocity $\overline{u_{\rm c}}^+$ is almost unchanged (or decreased $0.5\%$) 
with decreasing $Re_\tau$ from $80$ to $64$ in the case of XL, whereas increased by about $7\%$ 
between $Re_\tau=80$~(MB) and $64$~(LB), in which puff does not take place.

\Fref{fig:cf_nu} shows variations of the skin friction coefficient $C_f$ and the Nusselt number \Nus in comparison with the empirical correlations for a turbulent flow, together with the experimental data.
They are non-dimensionalized in the present definition as 
\begin{equation}
 C_f
 =\frac{\tau_\textrm{\tiny wall}}{\frac{1}{2}\rho u_\textrm{\tiny m}^2}
 =\frac{2}{{u_\textrm{\tiny m}^+}^2}, \hspace{1em}
 N\hspace{-0.15em}u
 =\frac{h \cdot 2\delta}{\lambda }
 =\frac{2Re_\tau Pr}{\theta_\textrm{\tiny m}^+},
\end{equation}
where $\tau_\textrm{\tiny wall}$, $\rho$, $\theta_{\rm m}$, $\lambda$ and $h$ are the averaged wall shear stress, the density, the bulk mean temperature, the thermal conductivity and the heat transfer coefficient, respectively.
The wall shear stress and  heat transfer are locally enhanced by the puff-like structure as indicated already. 
As a result, both of $C_f$ and $\mathit{Nu}$ with the puff tend to be slightly higher than those without puff, and stay closer to the empirical correlations even for very low Reynolds numbers.

\section{Conclusions}

We performed DNS of turbulent heat transfer in a transitional channel flow with two different thermal boundary conditions down to $Re_\tau=60$, and investigated the characteristics of the flow and thermal fields accompanied by a puff-like structure (called `puff', hereafter). 

The puff occurs over the Reynolds-number range $Re_\tau \leq 80$ ($Re_{\rm m}\leq 2300$, or $Re_{\rm c} \leq 1430$) with the computational domain of $(L_x \times L_z) = (51.2\delta \times 22.5\delta)$. 
The puff maintains the intermittent turbulence as low as $Re_\tau=60$ ($Re_{\rm m}= 1640$, or $Re_{\rm c} = 1070$), whereas the flow without the puff becomes laminar at the larger $Re$ ($Re_\tau = 60$--$64$) than the transitional Reynolds number obtained in experiments \cite{Patel69,Carlson82}.
The puff oblique with respect to $x$ and $z$ directions is observed to emerge from an initial random distribution, although the influence of the periodic boundary condition cannot be neglected.
The homogeneous direction of the puff is tilted at an angle of $\tan^{-1}(L_z/L_x)$ ($=24^\circ$ in this case) to the $x$ direction.
The turbulent and heat transfer properties of the puff are investigated. 
It is revealed that the secondary flow around the laminar-turbulent interface induces the isolated strong-turbulence region.
The streaky structure usually found in fully developed turbulent flows are also present in the intermittent-turbulence part of the puff.
In consequence, an localized large $C_f$ and \Nus region occurs: moreover, their ensemble-averaged values are significantly larger than those without puff for $Re_\tau \leq 64$.

\section*{Acknowledgements}

The author would like to thank Prof. Hiroshi Kawamura and Dr. Kaoru Iwamoto for fruitful discussions and careful reading of the manuscript.
The present computations were performed with use of the supercomputing resources at Cyberscience Center of Tohoku University. 

\vspace{1em}

This paper is a revised and expanded version of a paper entitled ``DNS of Heat Transfer in a Transitional Channel Flow Accompanied by a Turbulent Puff-like Structure'', presented by T. Tsukahara, K. Iwamoto, H. Kawamura, and T. Takeda, at the 5th Int. Symp. on Turbulence, Heat and Mass Transfer, Dubrovnik, Croatia, Sep. 25--29 (2006), pp. 193--196.



\begin{thebibliography}{21}


\bibitem{Kim89}
J.~Kim and P.~Moin, 
\newblock Transport of passive scalars in a turbulent channel flow. 
\newblock In {\em Turbulent Shear Flows 6 
{\rm (}Edited by J.-C.~Andr\'e{\rm )}}, pp.~85--96, Springer-Verlag, Berlin, 1989.

\bibitem{Lyons91a}
S.~L.~Lyons, T.~J.~Hanratty and J.~B.~Mclaughlin, 
\newblock Large-scale computer simulation of fully developed turbulent channel flow with heat transfer. 
\newblock {\em Int. J. Num. Method in Fluids}, 13: 999--1028, 1991.

\bibitem{Lyons91b}
S.~L.~Lyons, T.~J.~Hanratty and J.~B.~Mclaughlin, 
\newblock Direct numerical simulation of passive scalar heat transfer in a turbulent channel flow. 
\newblock {\em Int. J. Heat and Mass Transfer}, 39: 1149--1161, 1991.

\bibitem{Kasagi92}
N.~Kasagi, Y.~Tomita and A.~Kuroda, 
\newblock A direct numerical simulation for passive scalar field in a turbulent channel flows. 
\newblock {\em Trans. ASME {\rm C:} J. Heat Transfer}, 114: 598--606, 1992.

\bibitem{Kawa98}
H.~Kawamura, K.~Ohsaka, H.~Abe and K.~Yamamoto. 
\newblock DNS of turbulent heat transfer in channel flow with low to medium-high Prandtl number fluid.
\newblock {\em Int. J. Heat and Fluid Flow}, 19: 482--491, 1998.

\bibitem{Kawa00}
H.~Kawamura, H.~Abe and K.~Shingai. 
\newblock DNS of turbulence  and heat transport in a channel flow with different Reynolds and Prandtl numbers and boundary conditions.
\newblock In {\em Third International Symposium on Turbulence, Heat and Mass Transfer 
{\rm (}Edited by Y.~Nagano, K.~Hanjalic and T.~Tsuji{\rm)}},
pp.~15--32, 2000.

\bibitem{Iida98}
O.~Iida and Y.~Nagano. 
\newblock The relaminarization mechanisms of turbulent channel flow at low Reynolds numbers. 
\newblock {\em Flow, Turbulence and Combustion}, 60: 193--213, 1998.

\bibitem{Tsukahara05}
T.~Tsukahara, Y.~Seki, H.~Kawamura and D.~Tochio. 
\newblock DNS of turbulent channel flow at very low Reynolds numbers.
\newblock In {\em Fourth International Symposium on Turbulence and Shear Flow Phenomena 
{\rm (}Edited by J.~A.~C.~Humphrey et al.{\rm )}},
pp.~935--940, 2005: 
T.~Tsukahara. 
\newblock DNS of turbulent channel flow at very low Reynolds numbers.
\newblock {\em arXiv}. 1406.0248.

\bibitem{Wygnanski73}
I.~J.~Wygnanski and F.~H.~Champagne. 
\newblock On transition in a pipe. Part 1. The origin of puffs and slugs and the flow in a turbulent slug. 
\newblock {\em J. Fluid Mech.}, 59: 281--335, 1973.

\bibitem{Davies28}
S.~J.~Davies and C.~M.~White. 
\newblock An experimental study of the flow water pipes of rectangular section. 
\newblock {\em Proc. R. Soc. Lond.}, A 119: 92--107, 1928.

\bibitem{Patel69}
V.~C.~Patel and M.~R.~Head. 
\newblock Some observations on skin friction and velocity profiles in fully developed pipe and channel flows. 
\newblock {\em J. Fluid Mech.}, 38: 181--201, 1969.

\bibitem{Carlson82}
D.~R.~Carlson, S.~E.~Widnall and M.~F.~Peeters. 
\newblock A flow-visualization study of transition in plane Poiseuille flow. 
\newblock {\em J. Fluid Mech.}, 121: 487--505, 1982.

\bibitem{Orszag71}
S.~A.~Orszag. 
\newblock Accurate solution of the Orr Sommerfeld stability equation. 
\newblock {\em J. Fluid Mech.}, 50: 689--703, 1971.

\bibitem{Orszag80}
S.~A.~Orszag and L.~C.~Kells. 
\newblock Transition to turbulence in plane Poiseuille flow and plane Couette flow. 
\newblock {\em J. Fluid Mech.}, 96: 159--205, 1980.

\bibitem{Klingmann90}
B.~G.~B.~Klingmann and H.~Alfredsson. 
\newblock Turbulent spots in plane Poiseuille flow -- Measurements of the velocity field. 
\newblock {\em Phys. Fluids}, A2: 2183--2195, 1990.

\bibitem{Henningson91}
D.~S.~Henningson and J.~Kim. 
\newblock On turbulent spots in plane Poiseuille flow. 
\newblock {\em J. Fluid Mech.}, 228: 183--205, 1991.

\bibitem{Wygnanski75}
I.~J.~Wygnanski, M.~Sokolov and D.~Friedman. 
\newblock On transition in a pipe. Part 2. The equilibrium puff. 
\newblock {\em J. Fluid Mech.}, 69: 283--304, 1975.

\bibitem{Abe01}
H.~Abe, H.~Kawamura and Y.~Matsuo. 
\newblock Direct numerical simulation of a fully developed turbulent channel flow with respect to the Reynolds number dependence. 
\newblock {\em Trans. ASME {\rm I:} J. Fluids Engng.}, 123: 382--393, 2001.

\bibitem{Henningson87}
D.~S.~Henningson, P.~Spalart and J.~Kim. 
\newblock Numerical simulations of turbulent spots in plane Poiseuille and boundary-layer flow. 
\newblock {\em Phys. Fluids}, 30: 2914--2917, 1987.

\bibitem{Priymak04}
V.~G.~Priymak and T.~Miyazaki. 
\newblock Direct numerical simulation of equilibrium spatially localized structures in pipe flow. 
\newblock {\em Phys. Fluids}, 16: 4221--4234, 2004.

\bibitem{Tennekes72}
H.~Tennekes and J.~L.~Lumley. 
\newblock In {\em A first course in turbulence}, MIT Press, Cambridge, MA, 1972.

\bibitem{Dean78}
R.~D.~Dean. 
\newblock Reynolds number dependence of skin friction and other bulk flow variables in two-dimensional rectangular duct flow. 
\newblock {\em Trans. ASME {\rm I:} J. Fluids Engng.}, 100: 215--222, 1978.

\bibitem{Kays80}
W.~M.~Kays and M.~E.~Crawford. 
\newblock In {\em Convective heat and mass transfer}, 2nd ed., McGraw-Hill, New York, 1980.

\end{thebibliography}
\end{document}